\documentstyle[11pt]{article}
\newcommand{\be}{\begin{equation}}
\newcommand{\ee}{\end{equation}}
\newcommand{\pa}{\partial}

\begin{document}

\title{
Non-uniqueness of the third post-Newtonian binary point-mass dynamics}

\author{Piotr Jaranowski$^{1,2}$ and Gerhard Sch\"afer$^3$
\\{\it $^1$Institute of Physics, Bia{\l}ystok University}\\
  {\it Lipowa 41, 15-424 Bia{\l}ystok, Poland}
\thanks{E-mail: pio@alpha.fuwb.edu.pl}
\\{\it $^2$Albert-Einstein-Institut,
       Max-Planck-Institut f\"ur Gravitationsphysik}\\
  {\it Schlaatzweg 1, 14473 Potsdam, Germany}
\\{\it $^3$Theoretisch-Physikalisches Institut, 
       Friedrich-Schiller-Universit\"at}\\
  {\it Max-Wien-Platz 1, 07743 Jena, Germany}
\thanks{E-mail: gos@tpi.uni-jena.de}}

\date{}

\maketitle

\begin{abstract}

It is shown that the recently found non-uniqueness of the third post-Newtonian
binary point-mass ADM-Hamiltonian is related to the non-uniqueness at the third
post-Newtonian approximation of the applied ADM-coordinate conditions.

\vspace{0.5cm}\noindent PACS number(s):
04.25.Nx, 04.30.Db, 97.60.Jd, 97.60.Lf
\end{abstract}

\bigskip

In a recent paper \cite{JS97} the authors reported on the non-uniqueness of the
3rd post-Newtonian ADM-Hamiltonian for binary point-mass systems.  The term in
the Hamiltonian which came out to be ambiguous, in the center-of-mass reference
frame, is given by (see Eqs.\ (71) and (75) in \cite{JS97})
\be
\omega \frac{G^3 m_1 m_2}{c^6} p_{1i}p_{1j}
\pa_{1i}\pa_{1j}\left(-\frac{1}{r_{12}}\right).
\ee
In Eq.\ (1) $m_1$ and $m_2$ denote the masses of the bodies 1 and 2,
respectively, $r_{12}=|{\bf x}_1-{\bf x}_2|$ is their relative coordinate
distance, where ${\bf x}_a$ ($a=1,2$) denotes the position of the $a$th body.
For the momenta of the bodies $p_{1i} = - p_{2i}$ holds $(i=1,2,3)$; $\pa_{ai}$
denotes the partial derivative with respect to $x_a^i$.  $G$ and $c$ are the
Newtonian gravitational constant and the speed of light, respectively.  The
ambiguity in the Hamiltonian is expressed by an unspecified finite number
$\omega$.

In our treatment we applied the following generalized isotropic ADM-coordinate
conditions (see Eqs.\ (7-4.22) and (7-4.23) in \cite{ADM62}, respectively
Eqs.\ (3) and (4) in \cite{JS97})
\begin{eqnarray}
g_{ij}&=&\left(1+\frac{1}{8}\phi\right)^4\delta_{ij}+h^{TT}_{ij},
\\[2ex]
\pi^{ii}&=&0,
\end{eqnarray}
where $g_{ij}$ denotes the 3-metric and $\pi^{ij}$ the field momentum (canonical
conjugate to $g_{ij}$), $h^{TT}_{ij}$ is the transverse-traceless (with respect
to the flat-space metric) part of $g_{ij}-\delta_{ij}$.  The form of the
isotropic part of Eq.\ (2) stems from the Schwarzschild metric in isotropic
coordinates.

To leading order in powers of $1/c$ (this is enough for the following)
the coordinate condition (3) reads ($x^0=ct$)
\be
2 \pa_{i} g_{0i} - \pa_{0} g_{ii} = 0.
\ee
Note that the coordinate conditions (2) can exactly be written as
\be
3 \pa_{j} g_{ij} - \pa_{i} g_{jj} = 0.
\ee

Let us define the following infinitesimal coordinate transformation
\be
x'^{\mu} = x^{\mu} + \epsilon^{\mu}, \quad
\epsilon^0 = 0, \quad
\epsilon^i = \alpha \frac{G^3 m_1^2 m_2^2}{2c^6 M}
\pa_{i} (r^{-1}_{1} + r^{-1}_{2}),
\ee
where $M=m_1+m_2$ denotes the total mass of the system, $r_a=|{\bf x}-{\bf
x}_a|$, and $\alpha$ is a pure number.  This transformation induces the
following leading order transformation in the metric coefficients, keeping the
independent variables fixed (see, e.g., Eq.\ (16) in \cite{DS85})
\be
g'_{\mu\nu} (t, x^i, x^i_a(t), p_{ai}(t))
= g_{\mu\nu} (t, x^i, x^i_a(t), p_{ai}(t))
- g_{\mu\lambda} \pa_{\nu} \epsilon^{\lambda}
- g_{\nu\lambda} \pa_{\mu} \epsilon^{\lambda}.
\ee
Hereof the transformation of the Hamiltonian results, plugging the expression
(7) into the test-mass Hamiltonian (see, e.g., Eq.\ (5.2) in \cite{S85}) and
identifying in turn the test mass with the source masses.  This gives
\be
H' = H + \alpha \frac{G^3m_1m_2}{c^6} p_{1i} p_{1j} \pa_{1i} \pa_{1j}
\left(\frac{1}{r_{12}}\right).
\ee

Outside the mass points, the transformation (6) keeps invariant the equations
(2) and (3), respectively (5) and (4).  Towards spacelike infinity, the
perturbation (6) dies out very fast, like $1/r^2$, implying an $1/r^3$-decay
for the metric perturbation (7).

The shift in equation (8) is identical with equation (1).  This shows that the
dynamical ambiguity found in \cite{JS97} is related to the ambiguity in the
coordinate system (in quantum field theory those gauge ambiguities are well
known and result in the DeWitt-Faddeev-Popov ghost fields).  A corresponding
ambiguity is likely to exist also in harmonic coordinates as one can infer from
\cite{D83}, page 120.  For extended bodies, neither the coordinate-system
ambiguity arises nor is the dynamical ambiguity present.

\bigskip

\noindent
{\bf Acknowledgments}

\noindent This work was supported by the Max-Planck-Gesellschaft Grant No.\
02160-361-TG74 (GS) and by the KBN grant No.\ 2 P303D 021 11 (PJ).

\end{document}